\documentstyle{article}

\newlength{\defaultparindent}
\newenvironment{Default Paragraph Font}{}{}
\input{tcilatex}

\begin{document}

*;;;?;;Analysis of Quantum Evaporation Process of Black Holes

in the Model of Expansive Nondecelerative Universe\bigskip

Jozef Sima, Miroslav Sukenik

Slovak Technical University, Radlinskeho 9, 812 37 Bratislava, Slovakia

and Julius Vanko

Comenius University, Dep. Nucl. Physics, Mlynska dolina F1, 842 48
Bratislava, Slovakia

e-mail: sima@chelin.chtf.stuba.sk; vanko@fmph.uniba.sk\bigskip

Abstract. In the model of Expansive Nondecelerative Universe, black hole

cannot totally evaporate via quantum evaporation process proposed by
Hawking. In a limiting case, an equilibrium of gravitation creation and
black hole evaporation can be reached keeping the surface of its horizon

constant. This conclusion is in accordance with the second law of
thermodynamics.\bigskip

1. Introduction\medskip

One of the greatest achievements of cosmology and astrophysics was
formulation of hypothesis on the existence of black holes followed by
its
undirect experimental verification. Several problems relating to black
holes
have remained still open, one of them concerns ways of decreasing their
mass. Based on quantum mechanics and thermodynamics Hawking suggested a
solution of the problem in the form of quantum evaporation. His theory
has
led, however, to possibility of the total evaporation of black holes, a
phenomenon that has never been observed. In our previous contribution
[1] we
documented that a decrease in the mass of a black hole via its
evaporation
contradicts the second law of thermodynamics. In this contribution more
details are given and, in addition, independent modes of an evidence on
the
improbability of a black hole mass decreasing, based on mutual
consistency
of the calculations treated the gravitational field energy quantum,
output
and density, $E_{g},P_{g}$ and $\epsilon _{g}$ are offered.\bigskip

2. Theoretical background\medskip

In the model of Expansive Nondecelerative Universe (ENU) [1 - 3], the
gauge
factor $a$ , the cosmological time $t_{c}$ and the mass of the Universe
$%
M_{U}$ are related as follows\bigskip

$a=c.t_{c}=\frac{2G.M_{U}}{c^{2}}$ \hfill (1)\bigskip

Due to the matter creation, Vaidya metrics [4] is applied in ENU. This
mode
of treatment allows to localize the energy of the gravitational field.
Density of the gravitational field energy $\epsilon _{g}$ of a body with
the
mass $m$ in the distance $r$ is defined by relation\bigskip

$\epsilon _{g}=-\frac{R.c^{4}}{8\pi .G}=-\frac{3m.c^{2}}{4\pi .a.r^{2}}$

\hfill (2)\bigskip

where $R$ is the scalar curvature. Contrary to a more frequently used
Schwarzschild metrics (in which $\epsilon _{g}=0$ outside a body, and
$R=0$
), in Vaidya metrics $R\neq 0$ and $\epsilon _{g}$ may thus be
quantified
and localized also outside a body. For the energy of a quantum of the
gravitational field, equation (3) was derived in [1]\bigskip

$\left| E_{g}\right| =\left( \frac{m.\hbar ^{3}.c^{5}}{a.r^{2}}\right)
^{1/4} $ \hfill (3)\bigskip

Gravitational output, i.e. the amount of the gravitational energy
emitted by
a body with the mass $m$ within a time unit is defined as\bigskip

$P_{g}=\;\frac{d}{dt}\int \epsilon _{g}.dV\;\cong
\;-\frac{m.c^{3}}{a}\;=-%
\frac{m.c^{2}}{t_{c}}$ \hfill (4)\bigskip

Hawking [5] rationalized the process of quantum evaporation of a black
body;
within the process a black body with the diameter $r_{BH}$ evaporates
photons with the energy\bigskip

$E_{BH}=\frac{\hbar .c}{r_{BH}}$ \hfill (5)\bigskip

The output of a black hole evaporation was expressed by Hawking
as\bigskip

$P_{BH}=\frac{\hbar .c^{2}}{r_{BH}^{2}}$ \hfill (6)\bigskip

According to Hawking, a black hole with the mass $m_{BH}$ will totally
evaporate during the time\bigskip

$t\approx \frac{G^{2}.m_{BH}^{3}}{\hbar .c^{4}}$ \hfill (7)\bigskip

If the time $t$ is substituted by cosmological time $t_{c}$
being\bigskip

$t_{c}\cong 4.5x10^{17}s$ \hfill (8)\bigskip

then black holes completing their evaporation at present should have the

initial mass\bigskip

$m_{BH}^{o}\approx 10^{12}kg$ \hfill (9)\bigskip

It was admitted by Hawking himself that in spite of a great efford, no
such
an evaporation was experimentally observed.\bigskip

3. Evaporation of black holes analysed from the viewpoint of ENU\medskip

In the ENU model [1], the mutually related creation of matter and of
gravitational energy simultaneously occurs. The laws of energy
conservation
still hold since the energy of gravitational field is negative and,
consequently, the total energy of the Universe is thus exactly equal to
zero
[2, 5]. In case of black holes, the matter creation and evaporation are
competitive opposed processes. Since a magnitude of the surface of a
black
hole horizon is proportional to entropy and the second law of
thermodynamics
may not be violated, during the matter creation and evaporation the
surface
of black hole horizon must not decrease. In a limiting case, when
amounts of
the created gravitation and evaporated matter are just balanced, the
surface
of the black hole horizon is constant. For such a case three postulates
can
be formulated:

a)the energy of gravitational field quanta is identical to the energy of

photons emitted at evaporation,

b) gravitational output equals to output of the evaporation,

c)density of the energy of black hole radiation is equal to the density
of
gravitational energy.

Justification of the above postulates is verified below.

a)It follows from (3) and (5) that\bigskip

$\left( \frac{m_{BH}.\hbar ^{3}.c^{5}}{a.r_{BH}^{2}}\right) ^{1/4}\cong
\frac{\hbar .c}{r_{BH}}$ \hfill (10)\bigskip

Using (1), Hawking relation (7) is directly obtained from (10) providing

that time $t$ represents the cosmological time $t_{c} .$

From the viewpoint of ENU, relation (10) represents a limiting (i.e. the

lightest) black hole in a given cosmological time. Such a limiting black

hole may exist in cases when requirements on the equilibrium of creation
and
evaporation, and those stemming from the second law of thermodynamics
are
met.

b) At the same time it follows from relations (4) and (6) that (in
absolute
values)\bigskip

$\frac{m_{BH}.c^{2}}{t_{c}}\cong \frac{\hbar .c^{2}}{r_{BH}^{2}}$ \hfill

(11)\bigskip

Based on (11), Hawking relation (7) can again be derived, its
interpretation
being identical to that offered in a).

c) Using the Stefan-Boltzmann law and relation (2), it must hold in the
mentioned limiting case\bigskip

$\frac{3m_{BH}.c^{2}}{4\pi .a.r_{BH}^{2}}\cong \frac{4\sigma .T^{4}}{c}$

\hfill (12)\bigskip

Relation (12) can be simplified using formulas\bigskip

$m_{BH}=\frac{r_{BH}.c^{2}}{2G}$ \hfill (13)\bigskip

$t_{c}=\frac{a}{c}$ \hfill (14)\bigskip

$T\cong \frac{\hbar .c}{r_{BH}.k}$ \hfill (15)\bigskip

$I=\sigma .T^{4}\cong \frac{c.k^{4}T^{4}}{(\hbar .c)^{3}}$ \hfill
(16)\bigskip

where $I$ is the intensity of radiation, $k$ is the Boltzmann constant.
Based on (16), relation (17) can be derived\bigskip

$k^{4}\cong \sigma .\hbar ^{3}.c^{2}$ \hfill (17)\bigskip

Applying relations (13) - (17) to (12) again Hawking relation (7) is
obtained in the meaning of limiting black hole in a given cosmological
time.\bigskip

4. Conclusions\medskip

$\cdot $ $\cdot $ The present contribution offers an independent
derivation
of Hawking formula concerning the quantum evaporation of black holes.

$\cdot$ $\cdot$ None of the arguments used contradicts the validity of
the
second law of thermodynamics.

$\cdot$ $\cdot$ A mutually consistent determination of the gravitational

field energy quantum, output and density, $E_{g} ,P_{g} $ and $\epsilon
_{g}
$ can be taken as an evidence on the correct localization and
quantization
of gravitational energy.

$\cdot $ $\cdot $ The calculations are of approximative nature when
applying
in the domain of strong fields.\bigskip

References\medskip

1.Sima, J. , Sukenik, M.: Preprint: gr-qc 9903090

2. Skalsky, V., Sukenik, M.: Astrophys. Space Sci. 178 (1991) 169

3. Skalsky, V., Sukenik, M.: Astrophys. Space Sci. 181 (1991) 153

4. Vaidya, P.C.: Proc. Indian Acad. Sci. A33 (1951) 264

5. Hawking, W.S.: A Brief History of Time: From the Big Bang to Black
Holes,
Bantam Books, New York, p. 129

\end{document}